# Big Tree Viewer: A high-performance browser-based phylogenetic tree viewer with agentic AI integration

John B. Allard[1,2]* and Sudhir Kumar[1,2]

[1] Institute for Genomics and Evolutionary Medicine, Temple University, Philadelphia, PA 19122, USA

[2] Department of Biology, Temple University, Philadelphia, PA 19122, USA

* Corresponding author: john.allard@temple.edu

## Abstract

Phylogenetic trees represent evolutionary history visually, and the ability to display and explore them is a key function in research across biological disciplines. Very large gene and species trees are becoming common with the increasing availability of low cost DNA sequencing, creating performance and informativeness challenges for display software originally designed for phylogenies with a few hundred to a few thousand tips. We discuss these limitations and introduce a browser-based Big Tree Viewer (BTV) highly optimized for fast display and biologically informative exploration of phylogenetic trees containing even a million tips. BTV can automatically map binomial species names to the NCBI or Catalogue of Life taxonomies and display colored ribbons to indicate taxonomic groups for enhanced exploration. Even without taxonomic mapping, BTV automatically labels groups of neighboring tips that share the first word of their names, such as a genus or gene family name, providing readable large-scale landmarks when individual tip labels are too dense to read. BTV supports metadata display, visual customization, tanglegram-like tree comparisons, statistics, figure export, and multiple tree layouts, including a new spiral form to display enormously large trees. BTV is also designed to integrate with common agentic AI tools, allowing agents to display, style, annotate, and inspect trees at the user's request. The BTV web app is freely available at https://BigTreeViewer.net and runs entirely in the user's browser, with no data uploads to any central server. Desktop versions of the BTV app for macOS, Windows, and Linux are also provided for download.

## Introduction

Phylogenetic trees depict the inferred evolutionary history of species and genes and are used throughout the life sciences. A tree diagram brings branching relationships into a single view, allowing researchers to see the broad course of evolutionary history and recognize patterns. As sequencing costs fall and molecular datasets expand, trees with thousands of species are becoming more common, with some species phylogenies exceeding 100,000 tips (Carruthers et al. 2026; TimeTree 2026; J. B. Allard, unpublished data). Microbial genomics and pathogen surveillance also generate very large trees, including phylogenies with millions of viral sequences (Parks et al. 2022; Sanderson 2022). In addition, projects like TimeTree of Life aggregate published phylogenies into supertrees with hundreds of thousands of tips (Kumar et al. 2022; TimeTree 2026). Similarly, researchers in functional genomics can assemble very large multigene family trees, as the number of homologous sequences in public databases has increased (Langschied et al. 2024; Policarpo et al. 2024; Fogg et al. 2026).

In our experience, existing tree-viewing software implementations are either too slow to display many of these large trees and/or they lack features that enable more meaningful biological exploration (Sanderson 2022; Wang et al. 2025). Even when a tree could be displayed easily, we found readability and interpretation to be limited by the information about the taxa displayed at the tips. With a few hundred tips or more, labels on a single screen or page can become difficult to read, often necessitating manual labels indicating the placement of named clades, genes, and taxonomic groups. For trees with thousands of tips or more, manual practices become impractical, so there has been little incentive to even attempt to view such enormous trees in their entirety. Collapsed clades and subsets can be used in place of whole trees above a certain scale (Letunic and Bork 2024), but this simplification comes at the cost of losing the insights that might be gained from rapidly exploring an entire large phylogeny.

We developed Big Tree Viewer (BTV) to address both the performance and readability of large phylogenetic trees. BTV is a browser-based app optimized for rapid loading and smooth exploration of trees at any scale, including those with a million tips or more. To keep the full view informative when tip labels are illegible, it automatically displays genus labels derived from binomial species names when available. It can also map species names to the NCBI Taxonomy (Schoch et al. 2020) or Catalogue of Life (Hobern et al. 2021) and automatically display colored, labeled ribbons for taxonomic groups (see **Fig. 1A-B**). These labels provide recognizable groups that help users identify regions of the tree and choose where to examine it more closely. The same features are useful for smaller trees whose tip names cannot all fit legibly in a figure.

Our analyses showed rapid loading and smooth navigation in both rectangular and circular layouts across trees containing 1,000 to 1,000,000 tips (see **Fig. 5**).

BTV also provides metadata displays (e.g., **Fig. 2**), visual customization, a tree comparison tool with a fully navigable tanglegram-like display (e.g., **Fig. 3**), statistics, and high-quality image export. Researchers can switch among rectangular, circular, and spiral layouts along with cladogram and chronogram options. The web app runs within the user's browser, and desktop versions are available for macOS, Windows, and Linux. BTV also provides interfaces for AI agents that let users request help with displaying, styling, annotating, and inspecting trees. We demonstrate features of BTV using species and gene trees and compare loading and navigation performance across web-based viewers.

## Results

### Biological orientation across viewing scales

BTV is developed for rapid loading and smooth exploration of trees at any scale, including those with a million tips or more. It is programmed to automatically recognize binomial species names in tip labels and extract genus names (first word in the tip label). It identifies consecutive runs of tips sharing the first word in the tip label and displays legible labels for larger genera, with additional genera visible as the magnification increases (**Fig. 1A**). The same grouping mechanism can label contiguous runs sharing another first-token identifier, such as a strain or gene-family name.

Users can add richer context to their trees by automatically mapping tip labels to the NCBI Taxonomy or Catalogue of Life (**Fig. 1B**). The NCBI taxonomy provides names and classifications associated with public sequence records across cellular organisms and viruses (Schoch et al. 2020). Catalogue of Life integrates expert species checklists and extends coverage beyond organisms represented in sequence databases (Hobern et al. 2021). BTV uses these sources to construct labeled ribbons for taxonomic ranks, allowing users to choose the source and customize the appearance. The displayed taxonomic detail adjusts as the user changes the zoom by using the mouse wheel or keyboard controls (**Fig. 1C**). By default, BTV displays Linnaean ranks, and users can customize which additional ranks are shown.

These annotations retain biological context as, for example, users navigate from the vertebrate overview toward mammals and the neighborhood of *Homo sapiens* (**Fig. 1C**). Search accepts species names and mapped taxonomic groups and provides navigation to matching regions.

Taxonomy colors can also be applied to branches, or reserved for ribbons when branch color represents another variable.

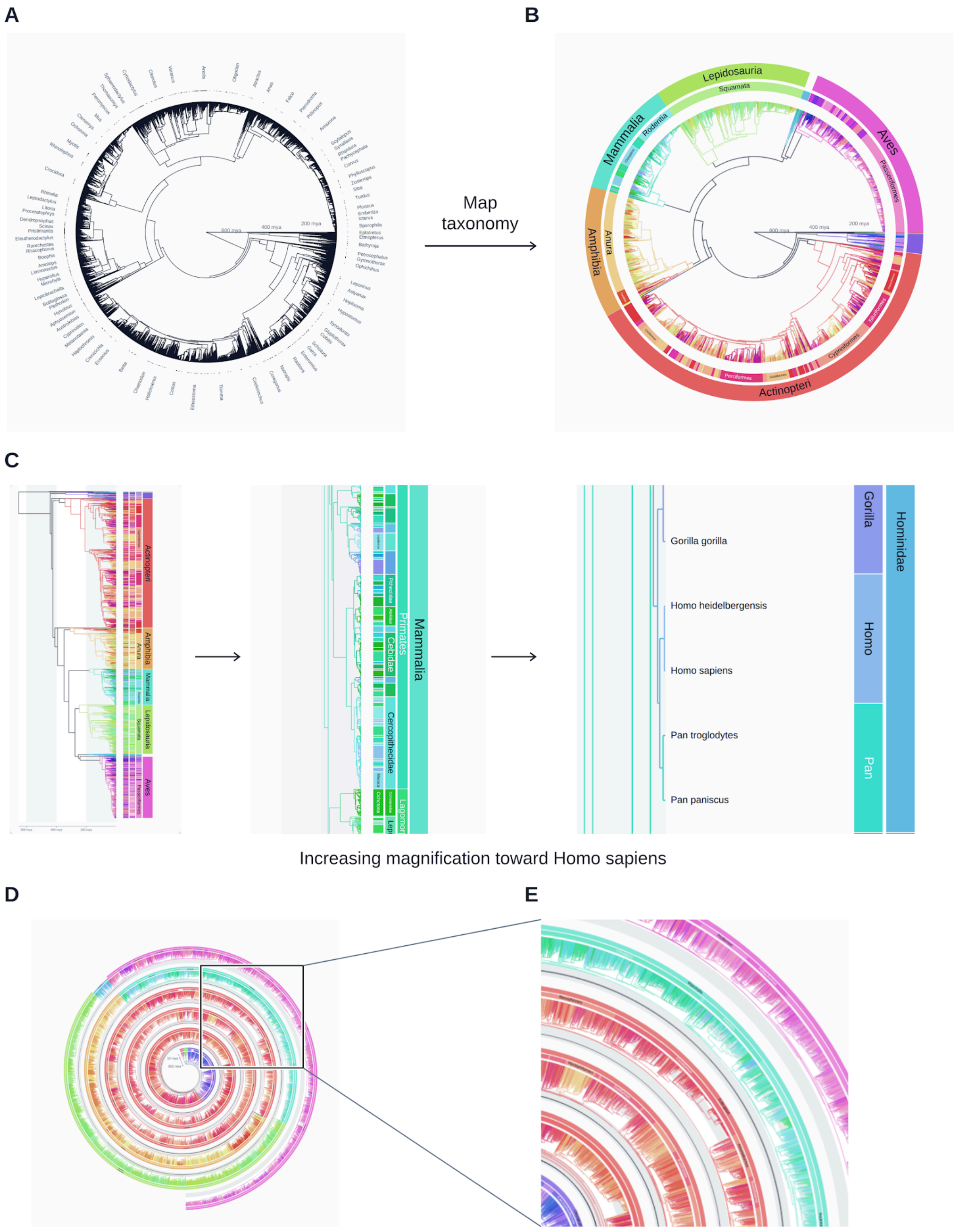

**Fig. 1. Taxonomy-guided exploration of a large phylogeny. (A-B)** The same circular view of a 50,033-tip vertebrate example tree before and after enabling NCBI taxonomy overlays. A displays automatic genus labels. In B, branch colors and nested ribbons identify taxonomic groups. **(C)** Successively higher magnification rectangular views, from the complete tree through mammals to the great-ape neighborhood. Arrows indicate successive views. **(D)** The complete tree in spiral layout. **(E)** Enlargement of the exact region boxed in D. Colors identify taxonomic groups.

## An interactive spiral representation of evolutionary time

BTV provides rectangular, circular, and spiral layouts for the same tree. The spiral is inspired by the 2015 TimeTree of Life, which arranged a logarithmically scaled evolutionary history into a coiled strip to display an extremely tall rectangular timetree of over 50,000 species (Hedges et al. 2015). In this representation, tip order proceeds along the strip and time progresses across its width. Position farther around the spiral does not mean a later evolutionary time. Compressing deep history while allocating more space to recent divergences creates a compact overview with a long boundary along which terminal groups and associated metadata can be displayed. Users can zoom into the spiral to inspect branches and labels and pan through neighboring groups, with taxonomy ribbons preserving group identity during navigation (**Fig. 1D-E**). This makes the layout useful for connecting broad metadata patterns with local relationships, particularly for quantities associated with extant species or recent branches.

## Metadata and figure preparation

In BTV, users can display metadata on the tree and customize the combined view for figure export. Metadata can be displayed through continuous and categorical branch colors, node markers, tip pie charts, heatmaps, bar charts, and tables (**Fig. 2**). Users can adjust fonts, label sizes, colors, line widths, ribbons, and annotation dimensions, and export PNG or SVG images. Saved sessions retain the tree and its configured display for later work.

The spiral geometry is particularly suitable for displaying metadata with a focus on more recent evolutionary patterns, such as speciation rates (**Fig. 2A**). Even with >50,000 tips in the spiral tree with metadata mapped across the branches, viewer navigation remains fast. A legend can be generated and displayed on the tree view and can be exported along with the tree image. BTV can display node names, heights, bootstrap values, and age error bars when these data are present in the loaded tree file. Users can customize these annotations together with branch-length and time-scale bars, ticks, and labels. Clades can be collapsed to triangular regions or assigned custom colors to draw focus to relevant regions as needed.

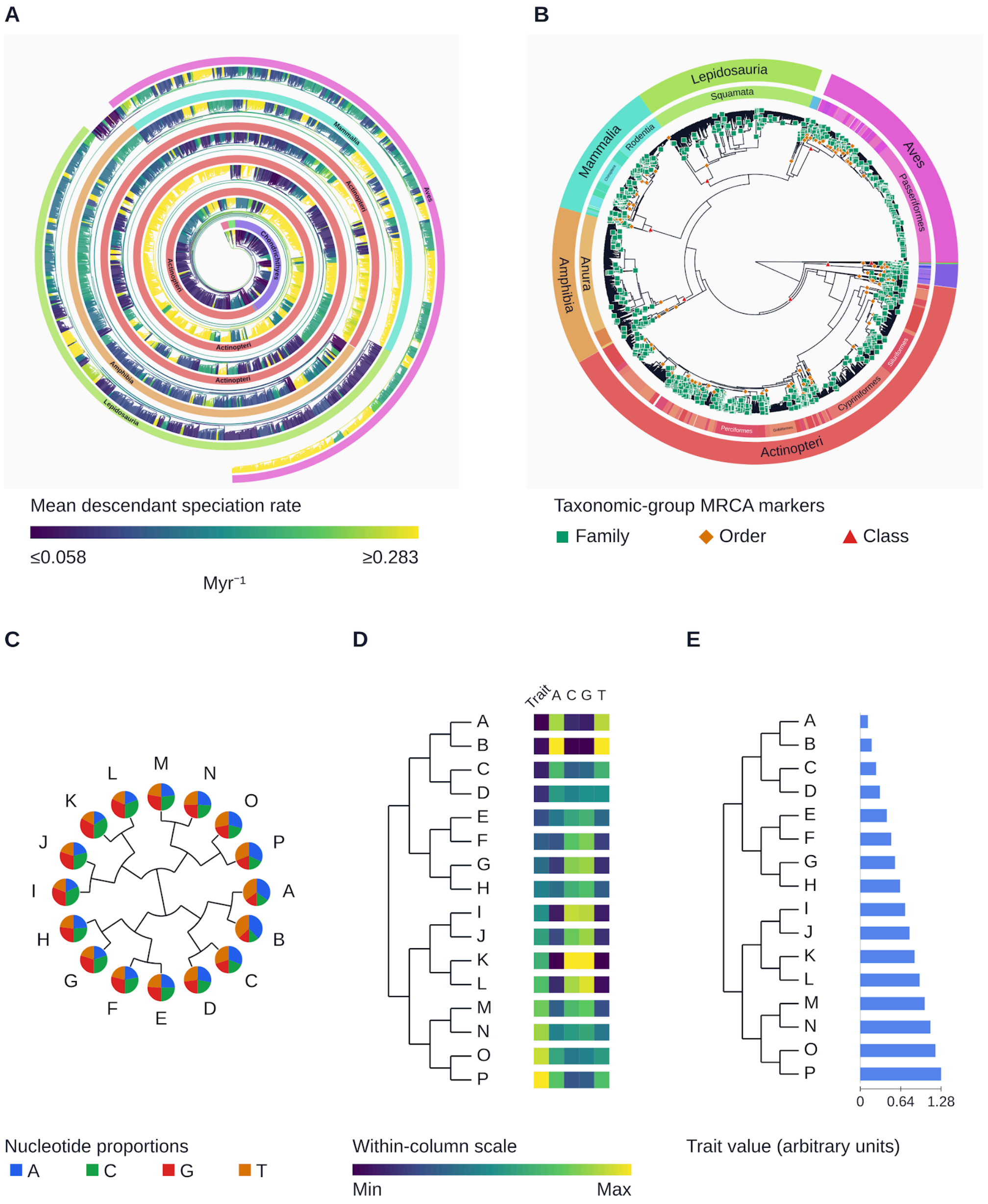

**Fig. 2. Continuous, categorical, and multivariate metadata displays. (A)** The 50,033-tip example vertebrate tree with mapped RPesto speciation rates. Internal colors summarize directly mapped descendant-tip rates. The scale is clipped at its first and ninety-ninth percentiles. Class ribbons encode taxonomy independently of branch color. **(B)** Markers at group most recent common ancestors: family, square; order, diamond; class, triangle. **(C-E)** The same illustrative 16-tip tree and metadata, with tip names abbreviated to A-P.

BTV also provides a navigable comparison of trees matched by shared tip labels (**Fig. 3**). Taxonomy ribbons retain biological context while users inspect differences in relationships and tip order. Connector crossings show display-order changes, while incompatible-split indicators flag topological disagreements.

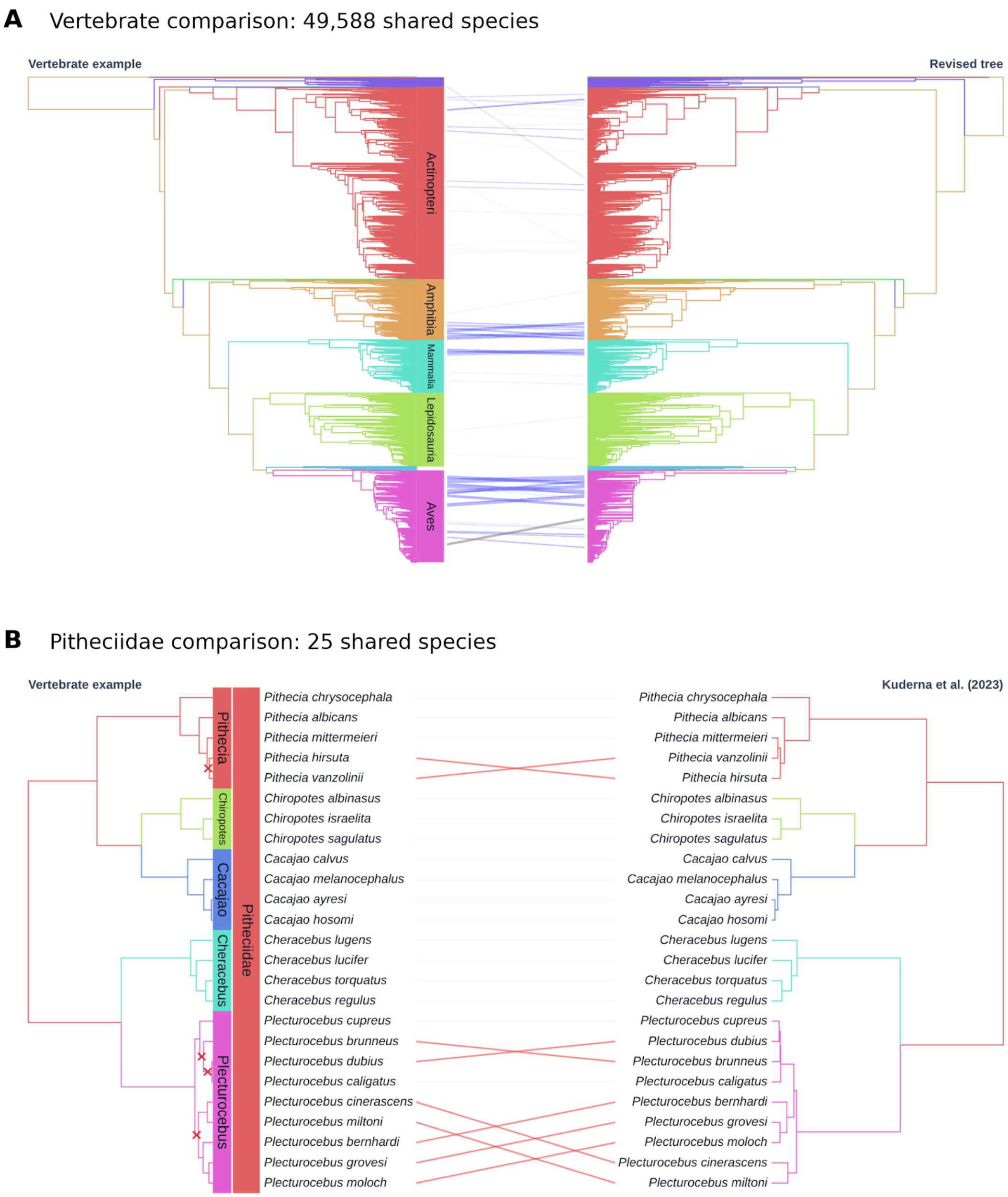

**Fig. 3. Tree comparison with taxonomic context.** (**A**) The vertebrate example and revised tree, restricted to 49,588 shared species, with class-level taxonomy colors and a ribbon. (**B**) The vertebrate example and the tree from Kuderna et al. (2023), restricted to 25 shared Pitheciidae species, with family and genus ribbons. Connectors join matching species; connector color emphasizes displacement in tip order, and red X marks indicate incompatible left-tree splits. B compares extracted subtrees. Each tree's horizontal depth is scaled independently.

For gene-family trees containing multiple tips per species spread across the phylogeny, taxonomy can be assigned from source-species names or NCBI taxonomy identifiers in the tip labels. Taxonomy and metadata can also be combined in large multi-gene trees, with outer ribbons identifying gene subfamilies and inner taxonomy ribbons identifying the source organisms (**Fig. 4A**). For samples collected at different times, BTV can display an already time-scaled phylogeny against calendar dates and color branches by date metadata, allowing historical samples to terminate before the most recent samples (**Fig. 4B**).

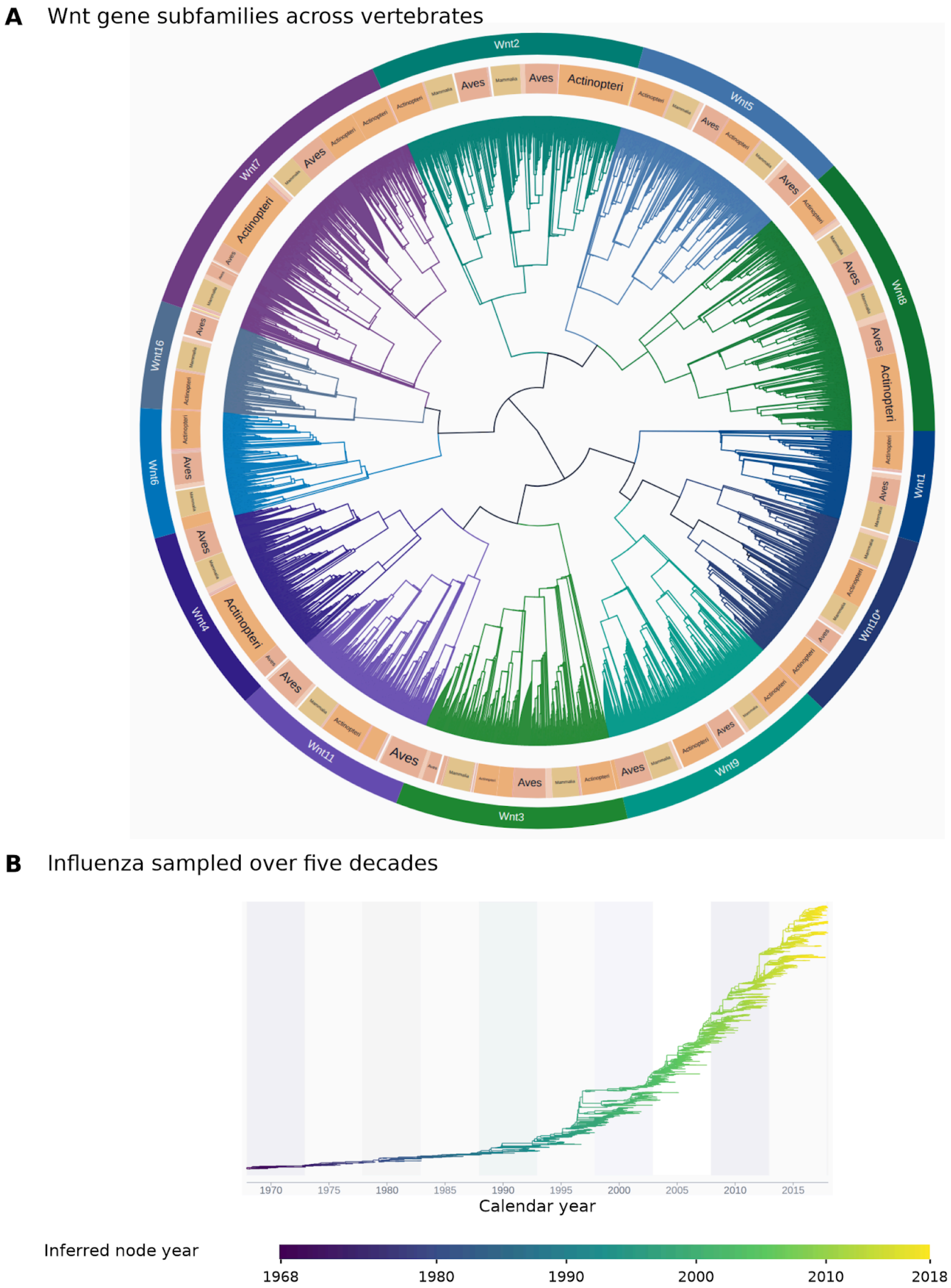

**Fig. 4. Gene-family and calendar-time displays.** (**A**) Published Wnt gene tree containing 38,886 sequences from 1,961 vertebrate species (Fogg et al. 2026). Branch colors and outer ribbons identify 12 Wnt subfamilies; the inner ribbon shows NCBI taxonomic classes mapped from source-organism names. The circular cladogram aligns the tips without representing genetic distance or evolutionary time. The asterisk marks a Wnt10 clade containing one sequence annotated as Wnt6 in the source data. (**B**) Published human influenza A/H3N2 haemagglutinin phylogeny containing 2,189 sampled viruses spanning 1968-2018 (Lee et al. 2018). Published inferred branch durations are displayed against calendar years, with branch colors depicting inferred node year. Historical samples terminate at their inferred dates rather than sharing a common present.

Users can share a saved session through a link to a hosted session file, allowing recipients to open the configured tree in their own browser. An embedding API is also provided so that a BTV viewer pane and tree session can be displayed inside a website, allowing visitors to explore a tree.

## Loading and navigation at increasing tree sizes

We compared the performance of BTV with that of iTOL (Letunic and Bork 2024), Taxonium (Sanderson 2022), PhyloScape (Wang et al. 2025), Phylo.io (Robinson et al. 2016), and phylotree.js (Shank et al. 2018), using identical balanced Newick trees across six sizes from 1,000 to 1,000,000 tips (**Fig. 5**). IcyTree (Vaughan 2017) and heat-tree (Foster et al. 2026) did not display a 100,000-tip tree within the 100 second screening limit and were, thus, excluded from the comparison. BTV, Taxonium, and PhyloScape displayed every tested size. BTV had the shortest median loading time at every tested size, ranging from 0.040 seconds for the 1,000-tip tree to 2.195 seconds for the 1,000,000-tip tree. Loading measured native initial displays rather than equal-density rendering: phylotree.js initially extended beyond the viewport, and iTOL and Phylo.io initially collapsed some clades.

To assess how smoothly the viewers displayed trees during navigation, we tested zooming and panning in rectangular and circular layouts. The test required zooming all the way in such that approximately ten tips were visible on the screen, followed by panning up and down. We measured the time between successive updates of the displayed tree and calculated the 95th percentile of these intervals to characterize the longer pauses. BTV outperformed all other tested viewers in both layouts for trees with over 1,000 tips. Taxonium and Phylo.io do not provide circular layouts in the tested applications. The hosted phylotree.js interface provides spacing controls, but its minimum rectangular tip spacing prevented fitting the complete tree into the benchmark viewport. Its circular controls could not enlarge the tree enough to show approximately ten tips, so it contributes loading results only.

iTOL loaded trees with up to 100,000 tips and could zoom in until approximately ten tips were visible in both rectangular and circular layouts, but navigation became less smooth as tree size increased. At 100,000 tips, its median trial-level 95th-percentile display intervals were 5.30 seconds in rectangular view and 5.91 seconds in circular view. PhyloScape's Large Tree mode displayed all tested sizes, with rectangular intervals of 46.4 milliseconds at 100,000 tips and 548 milliseconds in its single successful navigation trial at 1,000,000 tips. Three imports during navigation setup returned HTTP 504 errors, leaving two successful trials at 500,000 tips and one at 1,000,000 tips. PhyloScape's circular view imposed a maximum zoom level that increasingly restricted exploration as tree size grew. For the million-tip tree, even the closest view still covered about 5,000 tips. This zoom limit prevented completion of the navigation test for trees larger than 1,000 tips. We therefore omitted those timings and marked the 1,000-tip point with an asterisk in **Fig. 5C**. An initial screen of Evolview (Subramanian et al. 2019) encountered guest-service errors before submission, so it is not included in **Fig. 5.** Phylo.io could complete the rectangular zoom-and-pan test for trees with up to 100,000 tips, with a median trial-level 95th-percentile display interval of 5.44 seconds at that size. Its required Optimise text action was invoked at the transition from zoom to pan. At 50,000 tips, two of its three attempts passed final visibility verification; in the third, most adjacent tip names remained absent after 90 seconds.

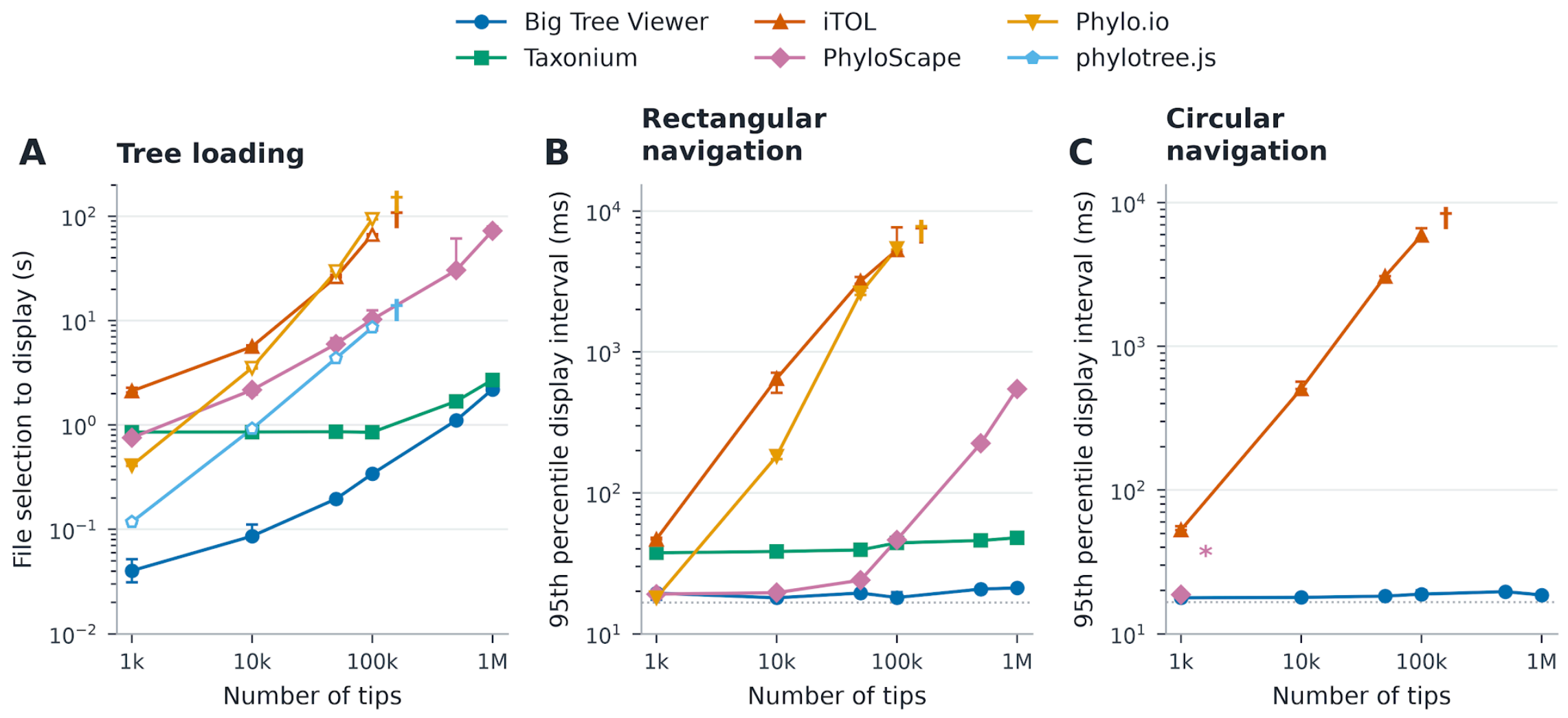

**Fig. 5. Tree loading and navigation responsiveness. (A)** Time from local-file selection to initial tree display, including required upload and server processing. Open symbols indicate initially collapsed clades (iTOL, Phylo.io) or a partial on-screen view (phylotree.js). **(B-C)** Trial-level 95th-percentile display intervals during rectangular (B) and circular (C) zoom-and-pan navigation from the full tree to a view of roughly ten adjacent tips. Longer intervals indicate larger pauses; the dotted reference is 16.7 ms (60 frames per second). Points show medians and whiskers show observed ranges across successful trials. † Larger tested inputs exceeded the 100 second loading cutoff or failed during rendering. * PhyloScape's maximum circular zoom limit prevented completion of the navigation test for trees larger than 1,000 tips; those timings are omitted. Other missing curves reflect unavailable layouts or unmet navigation requirements. See Methods for procedures, trial counts, failures, and viewer-specific limitations.

## Discussion

Large-tree visualization becomes more useful when the overview identifies recognizable biological groups rather than only a dense pattern of branches. BTV's automatic genus labels and taxonomy ribbons provide those landmarks, helping users decide which regions warrant closer inspection. Those landmarks also help researchers show where a biological pattern occurs in the tree. Users can locate a broad metadata pattern within a recognizable taxonomic group, zoom closer to identify the contributing species, and prepare an overview that retains those landmarks. The spiral extends the presentation introduced in the 2015 TimeTree of Life into a view whose local branches can be inspected interactively (Hedges et al. 2015). Shared sessions and embedded viewers could let readers examine relationships that a static overview cannot resolve.

The same approach can support very large multi-gene family trees, in which numerous gene copies from many organisms make individual tip labels impractical at overview scale. In the Wnt example, taxonomy ribbons identify the organisms represented within each gene-subfamily region, while separate outer ribbons distinguish the gene subfamilies (**Fig. 4A**). Distinct gene identifiers should be retained while organism names or taxonomic identifiers provide the annotation link. This application extends BTV beyond species phylogenies.

BTV is designed with agentic AI computing in mind. As science moves toward workflows assisted by AI agents, phylogenetic software needs interfaces that allow agents to inspect trees and help researchers display them. The desktop MCP tools, downloadable skill, and web API let agents configure trees for users and return editable sessions for interactive inspection. We intend to update these interfaces as agent capabilities evolve. The desktop application can check for and download software updates.

By retaining taxonomic context in large phylogenies, BTV can help researchers locate the lineages contributing to patterns in traits, evolutionary rates, and gene-family composition. As sequencing reveals evolutionary relationships at ever greater scales, these capabilities will help researchers uncover patterns, develop new hypotheses, and communicate findings across the full breadth of biodiversity.

## Materials and Methods

### Implementation

BTV is implemented in TypeScript and React, with an Electron desktop application sharing the web interface. A Web Worker parses the tree and constructs typed arrays for parent-child relationships, branch lengths, depths, and descendant counts. The worker precomputes input-and descendant-based leaf orders, including the range of descendant leaves under each node, and transfers the underlying buffers to the interface. This separates initial tree processing from the main interface thread and permits the reuse of topology and layout information during navigation.

The interactive renderer uses Canvas 2D. Layout-dependent geometry and indices are cached, with several caches constructed only when requested. Spatial indexing and viewport tests restrict detailed drawing to relevant segments. For uncolored rectangular trees and uncolored or large taxonomy-colored circular trees, the renderer traverses descendant intervals or angular sectors to exclude offscreen clades. Descendants spanning less than half a pixel are represented by their extent and expand into individual branches as the view enlarges. The same circular traversal applies across tree sizes; other annotated rendering paths retain their own caches and detail controls. Tip and group labels are selected according to available screen space, avoiding the cost and visual clutter of drawing every label in an overview. For large taxonomy-colored circular trees with the tree fully in view, panning reuses a branch image at the current scale while labels and ribbons continue to update for the viewport. Changing scale invalidates that image and retains the clade-based zoom renderer. Redraws are scheduled through requestAnimationFrame. Cached whole-tree paths are used selectively: large paths or deep zooms can instead use visible-segment rendering. The implementation therefore combines preprocessing, cache reuse, visibility filtering, and reduced detail where individual features cannot be resolved.

Taxonomy processing uses a separate worker and cached mappings. The renderer constructs taxonomic blocks from mapped tips in display order and reuses them as the camera changes. The spiral maps leaf positions along a coiled curve and converts age to a cross-strip offset using a normalized $log(1 + age/u)$ transformation, where $u$ controls the near-present scale. This transformation remains finite at age zero. For a dated ultrametric tree, *age* is calculated from *root age* minus root-to-node distance.

### AI-assisted tree exploration and figure preparation

The desktop application includes a local Model Context Protocol (MCP) server. Its tools let agents open Newick, NEXUS, and saved-session files in a visible viewer, inspect tree counts and annotation status, change layouts and display settings, and export figures or editable sessions. Agents can also render images in the background using the bundled Electron renderer and hand a configured tree to the researcher for continued interactive exploration. A single-command entry point exposes the same operations to scripts. The downloadable skill supplies instructions and helper scripts for choosing layouts, applying metadata and taxonomy, and checking rendered output. Agents can also use the web launch/export API to supply trees and display settings through URL parameters or structured browser messages and request images. The same API allows other websites to launch a configured tree for their users; it produced the class-ribbon spiral in **Fig. 2A** from a saved input payload. These interfaces reuse BTV's tree-loading and display operations, connecting automated figure preparation with interactive inspection.

## Example trees and metadata

**Fig. 1 and 2** use BTV's 50,033-tip vertebrate example session and its NCBI taxonomy mapping. The tree was inferred using divide-and-conquer phylogenetics as part of a forthcoming publication. Speciation-rate values from an existing RPesto analysis were used as illustrative branch-coloring metadata (Kopperud and Höhna 2025). Tip names were used to join the lambda estimates to the vertebrate tree. Values were available directly for 49,588 tips. Internal values were calculated as the mean among directly mapped descendant tips. Nodes without a finite value inherited the nearest ancestor's available value, producing 445 imputed tips and 140 imputed internal nodes. Class ribbons locate the rate patterns within recognizable biological groups while leaving the continuous branch-color scale intact.

The metadata table contains 100,065 unique node keys, all matched by the viewer. The continuous scale is clipped at its first and ninety-ninth percentiles, 0.0577383 and 0.2833427 $Myr^{-1}$. Class assignments were retained for 49,627 tips. The remaining 406 tips have no saved class assignment and appear as ribbon gaps. The source files, input checksums, mapping counts, display settings, and self-contained launch payload are provided in the Supplementary Data. The 16-tip examples use illustrative data to explain the visual encodings.

For **Fig. 3**, the revised vertebrate tree is a later version of the example tree. The overview compares the vertebrate example and revised tree on their 49,588 shared species; the detailed panel compares 25 shared Pitheciidae species from the vertebrate example and the tree of Kuderna et al. (2023). Tip labels were matched after replacing underscores with spaces and ignoring case. Unmatched species were removed, and unary nodes were suppressed while summing branch lengths. Taxonomy annotations were reused from the example's NCBI mapping. Each tree's horizontal depth was scaled independently.

The Wnt gene-family example (**Fig. 4A**) uses the published 38,886-tip tree and gene annotations from Fogg et al. (2026), available in Figshare

(https://doi.org/10.6084/m9.figshare.29516450.v1). Tip identifiers were reformatted to include source-organism names while retaining unique gene identifiers. Source annotations were grouped into 12 broad Wnt subfamilies for branch coloring and custom clade ribbons. NCBI taxonomy was mapped from organism names, with class-level ribbons displayed inside the subfamily ribbons. The 99 sequences from an octoploid hybrid could not be mapped and were left without taxonomy assignments. The tree was ordered smallest-first and displayed as a circular cladogram with aligned tips.

The calendar-time example (**Fig. 4B**) uses the published human influenza A/H3N2 haemagglutinin tree and inferred node dates distributed with Lee et al. (2018) at https://github.com/jbloomlab/Perth2009-DMS-Manuscript. We displayed this published time-scaled phylogeny against a calendar-year axis, with branches colored by the published node dates.

### Comparative performance protocol

Tests ran sequentially in isolated, hardware-accelerated headless Chrome 156.0.8078.47 sessions on an Intel Core i9-14900K workstation with 61.8 GB RAM and an NVIDIA GeForce RTX 4080 SUPER. Browser windows measured 1440 × 1000 CSS pixels at a device-pixel ratio of 1, with the same browser launch settings for all viewers. BTV source tag 0.1.25.1 was used for both loading and navigation. Tests used balanced, bifurcating, ultrametric Newick trees with no taxonomy or metadata overlays. The inputs, scripts, raw observations, and build and application-version records are provided in the Supplementary Data.

All loading measurements used the same browser-clock procedure, from the native file-change event to the matching compositor presentation of the rendered input tree. Required launch or upload actions were invoked as soon as enabled. Their preparation, upload, and server processing remained within the measured interval. Traces linked completed tree-drawing or SVG-update batches to their rendering frame, compositor activation, and presentation feedback. Input tip counts and rendered input labels were checked where available; nonblank tree images and unchanged screenshots separated by a 250 millisecond pause verified the display outside the timed interval. Each successful condition was measured in three fresh browser sessions, extended to five if the median of the first three was below one second. Testing stopped at the first confirmed loading failure or display exceeding 100 seconds. Loading measured each application's native initial display: iTOL initially used a circular view, whereas the other plotted viewers initially used rectangular views. Initial clade collapse and partial on-screen views are distinguished in **Fig. 5A**; these are not equal-density rendering workloads. PhyD3 (Kreft et al. 2017) was not timed because its hosted demo accepted phyloXML but failed the Newick control.

Before navigation, trees were fitted to a complete, nonblank, stable overview, and collapsed iTOL and Phylo.io clades were expanded. The hosted phylotree.js interface's minimum rectangular tip spacing was 2 pixels, preventing a full-tree overview at the tested sizes in our viewport; its circular controls could not zoom in far enough to show approximately ten tips, so it was excluded from navigation measurements. Loading during navigation setup used the same 100 second allowance. iTOL's native horizontal-scale control was set to tip count divided by 100 to retain a usable fitted aspect, with label size 12 and branch width 1 in both layouts. Each viewer was instructed to zoom from the full-tree view to a view showing approximately ten adjacent tips over four seconds. Magnification increased by the same proportion in equal time intervals. The view then moved vertically back and forth for two seconds, returning to its starting position. Movement extended 15% of the viewing-area height in each direction. View positions were scheduled by elapsed time, so viewers that fell behind skipped missed positions rather than replaying them. Slow rendering could extend the sequence beyond six seconds.

Navigation was attempted in three trials per loadable condition. Native Canvas 2D or WebGL drawing activity and SVG tree-group updates were matched through browser traces to compositor presentation feedback using the same rule for all viewers. Multiple updates reaching the same compositor frame were counted once. The fitted view provided the time-zero baseline; intervals included the initial wait for an updated frame. We calculated each trial's 95th-percentile display interval and summarized successful trials by their median and observed range. Three successful navigation trials contributed to each uncapped condition, except PhyloScape at 500,000 and 1,000,000 tips (two and one, respectively, after HTTP 504 import failures) and Phylo.io at 50,000 tips (two after a failed final visibility check). Frame intervals were calculated from presentation timestamps. All viewers had a 90 second allowance for final visibility verification. After navigation, the expected adjacent labels and branches had to appear in a composited screenshot and remain stable in another screenshot after a 250 millisecond pause. PhyloScape's native viewport callback and 500 millisecond deferred reset were retained, with final verification following that reset. Display intervals characterize update regularity; final branch and tip-label visibility was verified separately. All viewers used uniform magnification of both axes. The focal point moved linearly in screen coordinates from its fitted position toward approximately 80% of pane width and 50% of pane height, keeping the region of interest visible during zooming. Phylo.io's native Optimise text action was invoked once at the start of panning, and the resulting updates remained in the measured sequence. Native pane dimensions differed among viewers. Taxonium's minimap was disabled. PhyloScape imposed a maximum zoom level in circular view, limiting how closely large trees could be examined. For the million-tip tree, the closest view still covered about 5,000 tips. The zoom limit prevented completion of the navigation test for trees larger than 1,000 tips, so we omitted those timings. A single asterisk marks the largest tree for which PhyloScape completed the circular test (1,000 tips) in **Fig. 5C**.

## Software and data availability

Big Tree Viewer is available at https://bigtreeviewer.net/, with source code at https://github.com/John-Allard/big_tree_viewer. Benchmark inputs, scripts, raw results, build records, and figure source data are provided as supplementary data.

## Acknowledgements

This work was supported by the United States National Institutes of Health (R35GM139540 to S.K.) and the United States National Science Foundation (DBI-2505985 to S.K.).

OpenAI Codex was used through Codex CLI during initial development (recorded versions 0.110.0 through 0.142.1) and subsequently through Codex in the ChatGPT desktop application for Linux (multiple releases; version 26.1111.21308 at final manuscript preparation). It assisted with software development, testing and analyses, figure preparation, and manuscript editing. John B. Allard directed this work and made all major design decisions and scientific judgments. The authors extensively supervised and reviewed the AI-assisted work and take responsibility for the software, analyses, figures, and manuscript.